\documentclass{article}
\pdfoutput=1
\usepackage[T1]{fontenc} 
\usepackage[utf8]{inputenc} 
\usepackage{ismir,amsmath,cite,url}
\usepackage{graphicx}
\usepackage{color}

\usepackage{makecell, booktabs} 
\usepackage{arydshln} 

\usepackage{xcolor}

\usepackage{lineno}

\title{On-Line Audio-to-Lyrics Alignment Based on a \\ Reference Performance}

\multauthor
{Charles Brazier$^1$ \hspace{0.5cm} Gerhard Widmer$^{1,2}$} { 
$^1$Institute of Computational Perception, Johannes Kepler University Linz, Austria\\
$^2$LIT AI Lab, Linz Institute of Technology, Austria\\
{\tt\small {firstname.lastname}@jku.at}
}
\def\authorname{Charles Brazier, Gerhard Widmer}

\def\authorname{C. Brazier, and G. Widmer}

\begin{document}

\maketitle
\begin{abstract}
Audio-to-lyrics alignment has become an increasingly active research task in MIR, supported by the emergence of several open-source datasets of audio recordings with word-level lyrics annotations. However, there are still a number of open problems, such as a lack of robustness in the face of severe duration mismatches between audio and lyrics representation; a certain degree of language-specificity caused by acoustic differences across languages; and the fact that most successful methods in the field are not suited to work in real-time. Real-time lyrics alignment (tracking) would have many useful applications, such as fully automated subtitle display in live concerts and opera. In this work, we describe the first real-time-capable audio-to-lyrics alignment pipeline that is able to robustly track the lyrics of different languages, without additional language information. The proposed model predicts, for each audio frame, a probability vector over (European) phoneme classes, using a very small temporal context, and aligns this vector with a phoneme posteriogram matrix computed beforehand from another recording of the same work, which serves as a reference and a proxy to the written-out lyrics. We evaluate our system's tracking accuracy on the challenging genre of classical opera. Finally, robustness to out-of-training languages is demonstrated in an experiment on Jingju (Beijing opera).
\end{abstract}

\section{Introduction}\label{sec:introduction}

Audio-to-lyrics alignment aims at synchronizing an audio recording with its corresponding lyrics, in order to retrieve the position of spoken or sung textual units in the recording. The task has been widely researched in the context of speech data \cite{McAuliffe2017, kida2021label}, and recently there has also been very promising work on polyphonic music \cite{dzhambazov2014automatic, stoller2019end, gupta2020automatic, demirel2021low}, even on multilingual alignment in a single framework \cite{vaglio2020multilingual}. Robust alignment methods would be useful for applications such as automatic karaoke captioning \cite{kan2008lyrically}, music or video cutting based on the lyrics, or automatic subtitling in music videos. Other tasks in Music Information Retrieval (MIR), such as cover detection or score following, could also benefit.

All proposed audio-to-lyrics alignment methods are composed of an acoustic model, classifying each audio frame into a set of textual units, and an alignment procedure to obtain the desired lyrics timings. Previous works in the field \cite{vaglio2020multilingual, demirel2021low} use source separation systems as a pre-processing step to extract the singing vocals beforehand, even if in \cite{gupta2020automatic}, the authors mention that the vocal extraction algorithms can add artifacts in the vocals. In \cite{gupta2020automatic}, the authors improved their aligners by modeling vowel durations in their lexicons \cite{gupta2018automatic}, which permits taking into account certain pronunciation aspects. Also, in \cite{demirel2021low}, the alignment is done in several passes, to first spot keyword positions in the audio, and then consider several smaller alignments in between the keywords.

The challenging question of real-time audio-to-lyrics alignment remains open and has not yet been tackled in the literature. This type of application would have great value, especially in live concerts and operas where fully automated subtitle displays could directly help the audience in following the live story. However, due to the architecture of existing acoustic models, the types of alignment algorithms conventionally used, and the additional steps detailed above, previous methods are not suited to work in real-time.

In this work, we propose a first audio-to-lyrics alignment pipeline that can operate in real-time,\footnote{We will not actually measure runtimes in this paper; the important aspect of our method is that it solves the problem in an on-line fashion, without access to future information, and that we can quantify its theoretical latency, based on how it processes the input data.} in a language-independent way. Instead of using a pronunciation dictionary to translate the lyrics into phonemes, we propose to use another recording of the target piece as a reference and proxy to the lyrics.\footnote{Of course, this will only be practicable in certain domains, where reference recordings are available.} This method has been widely used in the domain of score following, for robust tracking during live orchestra \cite{arzt2015real} or opera \cite{brazier2020towards} performances. To this end, we first design an acoustic model that predicts a frame-wise probability distribution over a pre-defined set of phonemes. Each prediction is based on a very limited temporal window, using a future context of 280~ms which defines the delay of our system. Then, saving all the predictions in a posteriogram matrix, we perform an OnLine Time Warping (OLTW) alignment between this (incrementally computed) posteriogram and the one from another performance that has been generated beforehand.

In this paper, after presenting acoustic models and alignment strategies of existing works in Section~\ref{sec:related_works}, we will present our on-line alignment system in Section~\ref{sec:system}. The robustness and accuracy of our tracker will be evaluated on two distinct datasets of `art' music (opera) to be described in Section~\ref{sec:data_description}. Results and a discussion will be given in Section~\ref{sec:experiments}. Finally, our conclusions and open questions will be presented in Section~\ref{sec:conclusion}.

\section{Related work}\label{sec:related_works}




Audio-to-lyrics alignment is an active research topic that is constantly stimulated by a yearly musical challenge\footnote{\url{https://www.music-ir.org/mirex/wiki/2020:Automatic_Lyrics-to-Audio_Alignment_Results}} and the appearance of new open-source training datasets such as DALI \cite{meseguer2018dali} or DAMP \cite{smule2018damp}. Recent works follow a common pipeline. First, an acoustic model is trained to extract from the audio signal a \textit{`posteriogram'} that represents the frame-wise probability distribution over textual units through time. On the lyrics side, the text is first translated into a sequence of textual units that correspond to the classes of the trained acoustic model. Finally, an alignment algorithm is applied between the posteriogram and the lyrics' representation, to retrieve textual unit timings in the audio. In this section, we present existing acoustic models, detail the alignment process, and show their limitations to operate in real-time.

\subsection{Acoustic Model}\label{sec:acoustic_model}

Acoustic models are Deep Neural Networks whose task is to classify audio inputs into a sequence of probabilities over a set of predefined textual units representing the lyrics present in the audio. They are trained with datasets that include a multi-modal mapping between lyrics and audio. Due to the difficulty (or near impossibility) of obtaining ground truth frame-level annotations, acoustic models are generally trained with weak annotations, at the sentence or word level, where the precise alignment between audio frames and lyrics remains unknown. Inspired by Speech Recognition \cite{sak2015learning}, models can be trained in different ways. A first strategy, used in \cite{gupta2020automatic, demirel2021low}, fits a Gaussian Mixture Model Hidden Markov Model (GMM-HMM) that force-aligns the lyrics with the audio to generate phone labels at the frame level. Then, the acoustic model is trained at the sequence level with the Lattice-Free-Maximum Mutual Information (LF-MMI) loss function \cite{povey2016purely}, considering the output of the GMM-HMM as ground truth. \cite{demirel2021low} combine this objective with the Cross-Entropy (CE) loss to train the model at the frame level. Another strategy, used in \cite{stoller2019end, vaglio2020multilingual}, aims at directly aligning the audio with the lyrics, using the \textit{Connectionist Temporal Classification (CTC) loss} \cite{graves2006connectionist}.

Acoustic models classify each audio frame into a set of textual units, which are intermediate lyrics representations. In \cite{stoller2019end}, the lyrics are represented as a sequence of characters, whereas \cite{gupta2020automatic, vaglio2020multilingual, demirel2021low} use a phoneme representation.
In the case of a single multilingual acoustic model, using a character representation is delicate. Even within one language, a letter can be pronounced differently depending on the context, which can confuse the acoustic model that tries to classify audio frames into letters. The phoneme representation is more consistent across different languages and provides better performance since it is not language-specific.
In \cite{vaglio2020multilingual}, the authors report better results in using phonemes as the intermediate representation.

Acoustic models can employ different network architectures. Existing architectures have been designed to take advantage of future information to improve the prediction at each time step, which limits their use to offline applications. In \cite{stoller2019end}, the authors build a Wave-U-Net that takes as input windows of raw audio and encodes the information at different scales. \cite{gupta2020automatic, demirel2021low} employ a combination of Convolutional Neural Networks and Time Delay Neural Network \cite{peddinti2015time} (TDNN-F) layers to model long temporal context. \cite{demirel2021low}  add CNN layers at the beginning of their model to speed up the training, and a multi-head attention layer at the end to help focusing on different parts of the input for each prediction. Each layer is also responsible to extend the scope of input frames that have a direct influence on the output frame prediction, that is, its Receptive Field (RF). From the descriptions of these model architectures, we derive RF values higher than 1.5s\footnote{This rough calculation is only based on the respective stack of TDNN layers and is much higher in practice. For a full description of their architectures, we refer the reader to the scripts available at respectively \url{https://github.com/chitralekha18/AutoLyrixAlign} and \url{https://github.com/emirdemirel/ALTA}.}, which is not suitable for a real-time application. Finally, the authors in \cite{vaglio2020multilingual} use a Bidirectional Long Short-Term Memory (BLSTM) to model the temporal dependencies, which combines backward and forward information about the sequence for every prediction. Another downside of this architecture is that it is much slower to train \cite{liu2019time}.

\subsection{Alignment}\label{sec:alignment}

In the next step, an alignment algorithm is applied between the posteriogram generated by the acoustic model and the lyrics. To compare the two modalities, the lyrics must first be translated into a sequence of textual units matching with the classes of the trained acoustic model. This is generally done by using open-source pronunciation dictionaries such as CMUdict\footnote{https://github.com/cmusphinx/cmudict}, for English only, or Phonemizer\footnote{https://github.com/bootphon/phonemizer}, which covers several languages. Then, considering the posteriogram as our observation sequence and the target lyrics, Viterbi-based forced alignment is applied to find the most probable path in the posteriogram that generates the lyrics. In \cite{kruspe2017lyrics}, the author compares two trackers, one based on Dynamic Time Warping (DTW) between posteriograms and binary posteriograms generated from the lyrics; and one based on the Viterbi algorithm between the decoded lyrics from the posteriograms and their ground truth. The author reports that the first approach, analogous to the audio-to-midi alignment task, does not perform as well as the Viterbi-based method.

The alignment used in all the above works suffers from two main limitations. First, the alignment algorithm works on the full audio recording and the entire lyrics, which does not permit real-time or on-line application. Second, the phoneme-based approaches that yield the best alignment accuracies \cite{vaglio2020multilingual} are dependent on text-to-phoneme tools to translate the lyrics into a phoneme sequence, according to the corresponding language. This limits the scope to languages that are covered by these tools.

\section{Proposed system}\label{sec:system}

\begin{figure}[t]
\centering
\includegraphics[width=0.9\columnwidth]{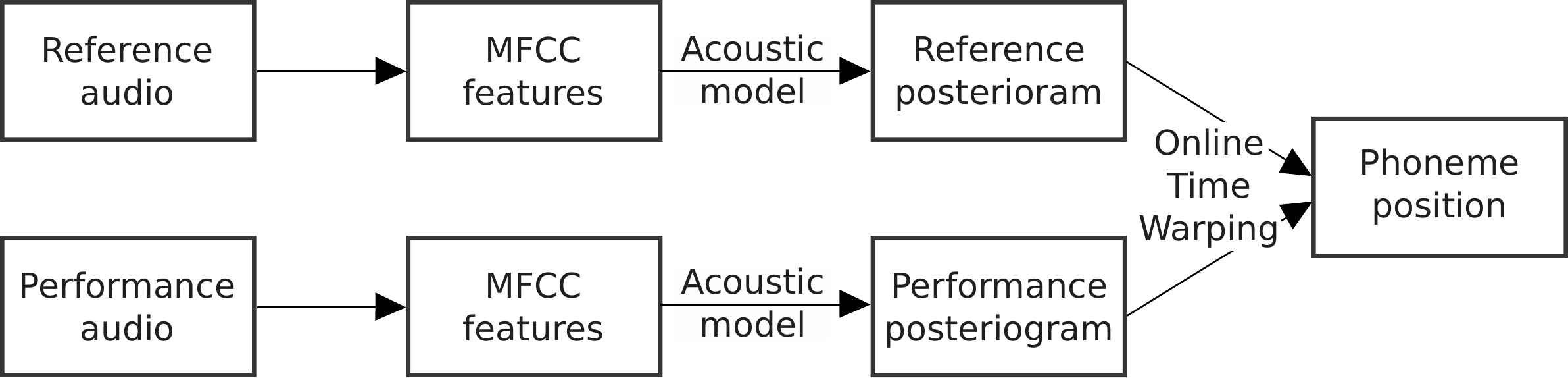}
\caption{Integrated Model.\label{fig:model}}
\end{figure}

Our proposed on-line audio-to-lyrics alignment system is illustrated in Figure~\ref{fig:model}. It is composed of an acoustic model that classifies in real-time each audio frame into a set of predefined phonemes. Then, instead of using the lyrics sequence for the alignment, we use a posteriogram matrix that is computed beforehand from another recording of the same work, which serves as a reference and a proxy to the written-out lyrics. Finally, we apply an OLTW alignment algorithm to align the two posteriograms, which permits to retrieve the position of the lyrics in the live recording with the help of manual lyrics annotations affixed to the reference.

\subsection{Acoustic Model}\label{sec:realtime_AM}

For our acoustic model, we select the CP-ResNet \cite{koutini2019receptive} architecture that has already proven to perform well in Acoustic Scene Classification \cite{koutini2020low} and in Emotion and Theme Recognition in Music \cite{koutini2019emotion}. Based on the ResNet architecture, the model stacks convolutional layers with additional residual connections between layers. The CP-ResNet is designed in such a way that the maximum time \textit{receptive field (RF)} is controlled by a hyper-parameter $\rho_{t}$ that defines the architecture of the model. For our experiments, we fix $\rho_{t}=6$, and our network architecture is given in Table~\ref{tab:architecture}. The RF of the model can be recursively calculated with stride and kernel size of each layer (see equation (1) in \cite{koutini2019receptive}). The corresponding RF is equal to 57~frames. This means that each output vector is dependent on 57~input frames centered around its time position, defining the latency of our model to 28~frames. The architecture of the deep network is specified in Table~\ref{tab:architecture}.

\begin{table}[t]
\small
 \begin{center}
 \scalebox{0.9}{ 
 \begin{tabular}{@{}ccccc@{}}
  \toprule
  \thead{\textbf{{\normalsize Layer}}}
    &   \thead{\textbf{{\normalsize Filters}}}
    &   \thead{\textbf{{\normalsize Kernel}}}
    &   \thead{\textbf{{\normalsize Stride}}}
    &   \thead{\textbf{{\normalsize Pad}}}\\
  \midrule
  \textbf{Conv2d+BN+ReLU} & 64 & 5$\times$5 & 2,2 & 1,1\\
  \textbf{Conv2d+BN+ReLU} & 64 & 3$\times$3 & 1,1 & 1,1\\
  \textbf{Conv2d+BN+ReLU} & 64 & 1$\times$1 & 1,1 & 1,1\\
  \textbf{MaxPool2d} & 1 & 2$\times$2 & 2,2 & 0\\
  \textbf{Conv2d+BN+ReLU (x6)} & 64 & 3$\times$3 & 1,1 & 1,1\\
  \textbf{Conv2d+BN+ReLU (x2)} & 128 & 3$\times$1 & 1,1 & 1,0\\
  \textbf{Conv2d+BN+ReLU (x2)} & 128 & 1$\times$1 & 1,1 & 0,0\\
  \textbf{Conv2d+BN+ReLU} & 60 & 1$\times$1 & 1,1 & 0,0\\
  \textbf{LogSoftmax} & - & - & - & -\\
  \bottomrule
 \end{tabular}}
\end{center}
 \caption{CP-Resnet with $\rho_{t}=6$.}
 \label{tab:architecture}
\end{table}

The acoustic model takes as input 80 Mel-Frequency Cepstral Coefficients (MFCC) features extracted from the audio signal, with a sampling rate of 16~kHz, and computed with a window size of 20~ms and a hop size of 10~ms. It corresponds to a model latency equal to 280~ms, which we consider acceptable for real-time applications such as, e.g., opera subtitling.

The model outputs a vector every 40~ms. The vector is of length 60 and includes 57~phonemes representing the union of all phonemes present in the English, German, French, Spanish and Italian languages, the space token, the instrumental token, and the mandatory blank token for CTC training. The five languages correspond to the most dominant languages present in the DALI dataset \cite{meseguer2018dali}, with a bias towards English: the dataset includes 225~hours of English songs, 20~hours of German, 10~hours of French, 10~hours of Spanish, and 10~hours of Italian, for a total of 275~hours with hierarchical annotations at the sentence, word, or note level. The dataset only covers Western musical genres. The choice of phoneme representation has been motivated by the multilingual aspect of our work, permitting to train a single model on different languages. The instrumental token proves to be useful to label audio inputs that do not contain singing voice, especially during silence or instrumental passages.

The model is trained on the DALI 1.0 \cite{meseguer2018dali} dataset with a CTC objective, which permits us to train a model with weak annotations between audio and lyrics. Each song in the dataset is cut into windows of 20~seconds with a hop size of 10~seconds, which limits the size of the input audio feature sequence to a maximum length of 2000~frames. Due to the real-time constraints, we do not extract the vocals from the audio mixture but we train our model with original mixtures of singing vocals and polyphonic music. The corresponding labels of each window are extracted with the word-level annotations provided by the dataset. Each word starting and ending in the corresponding time interval is part of the target annotation sequence of the corresponding audio window. Then, the character sequence is transformed into a phoneme sequence using Phonemizer, specifying the correct language (which is known at training time). Empty sequences are classified with the instrumental token. During training, the weights of the model are tuned to maximize the probability of getting the correct label sequence (or all derivative sequences that have inserted repetitions or blank symbols), given the input feature sequence.

\subsection{Alignment}\label{sec:realtime_Alignment}

The real-time alignment is realized between two audio performances of the same work, both containing the same sung lyrics. One performance, the \textit{target}, corresponds to the live performance we want to align to the lyrics. The other performance, the \textit{reference}, serves as a proxy to the written-out lyrics. This strategy has two main advantages. First, it is not dependent on a text-to-phoneme tool anymore. The pronunciation of the lyrics is contained in the reference recording and thus, can be decoded by the acoustic model. Even if the language of the target song is not included in the training set, the acoustic model maps both reference and target into the European phoneme set used during training. Also, the duration of the phoneme units is implicitly included in the reference recording, which means we do not need complex, explicit duration modeling techniques \cite{gupta2018automatic}. As a consequence, we can expect the reference posteriogram to be more similar to the posteriogram of the target recording we want to align to the lyrics. However, the reference has to be linked to the lyrics beforehand, generally with manual annotations at the word or sentence level -- but this really only depends on the requirements of the specific application.

We use the OLTW \cite{dixon2005line} algorithm to align reference and target posteriograms, skipping blank tokens in both reference and target sequence. The reference posteriogram is generated beforehand, feeding only the reference audio feature sequence to the acoustic model. Then, we generate in real-time, also with our acoustic model, the probability vectors representing the target posteriogram.
For each new vector, we calculate its cosine distance with a range of 8000 posteriogram vectors, centered around the expected position in the reference posteriogram.
This corresponds to a context of 320~seconds. Then, we calculate recursively the global cost, applying the standard DTW formula (equation (4.5) in \cite{muller2007dynamic}). The index representing the minimum of the global cost represents the current time position in the reference posteriogram.

\section{Data description}\label{sec:data_description}

In this work, we select opera recordings to evaluate our system, for three reasons. First, live opera would be a direct beneficiary of this tracking method, which would support a fully automatic subtitle display in the opera house (or in a live streaming application). Secondly, opera lyrics are challenging to track. Indeed, the genre of classical music has been considered by far as the least intelligible genre among eleven other genres \cite{condit2015catching}. Thus, evaluating our system on opera data is a good robustness indicator. Finally, opera is a musical genre that consistently produces identical works in several copies, with the change of the entire set of artists. Popular datasets in audio-to-lyrics alignment such as Hansen\cite{hansen2012recognition}, Mauch \cite{mauch2010lyrics} and Jamendo \cite{stoller2019end} do not include duplicate entries. The two opera datasets currently in our possession are described in Table~\ref{tab:datasets}.

\begin{table}[t]
\small
 \begin{center}
 \scalebox{1}{ 
 \begin{tabular}{@{}llcc@{}}
  \toprule
  \textbf{Opera} &  \textbf{Name} & \textbf{Duration} & \textbf{\# Annot.}\\
  \midrule
  \textbf{Don Giovanni} & Ref\_Karajan & 0:30:03 & 639\\
              & Targ\_Fischer & 0:34:58 & 639\\
              & Targ\_Manacorda & 0:30:40 & 639\\
  \midrule
  \textbf{Jingju} & Ref\_Jingju & 1:53:53 & 3,975\\
                  & Targ\_Jingju & 3:22:03 & 9,567\\
  \bottomrule
 \end{tabular}}
\end{center}
 \caption{Description of Don Giovanni and Jingju datasets.}
 \label{tab:datasets}
\end{table}

The first is a subset of the Italian opera \textit{Don Giovanni} by W.A.Mozart that covers all the \textit{recitativo} sections. These have been manually annotated with bar lines, making it possible for us to test the lyrics tracker by measuring how precisely it aligns target and reference at bar boundaries. Recitatives, an essential opera component of that period, have recently been in the focus of opera score following research  \cite{brazier2020towards}, but trackers remain brittle. This is due to the liberty that singers can take in terms of timing, singing style, etc., and the fact that musical accompaniment is often improvised and played by different instruments in different recordings. Thus, it would be helpful to be able to follow the performance based on the content of the sung lyrics. As reference and proxy to the lyrics, we use a CD recording conducted by Herbert von Karajan in 1985. The two live targets are performances that were recorded at the
Vienna State Opera
in 2018 and 2019 and conducted, respectively, by \'Adam~Fischer and Antonello~Manacorda, with completely different casts of singers. For each performance, the complete subsections comprising the recitativo sections only, contain 639 manual bar-level annotations for a duration of approximately 30~minutes.

The second dataset we will use is a subset of the Jingju (Beijing Opera) A Capella Singing Audio Dataset \cite{caro2014creating, gong2017creating}. It has been recorded in a teacher/student manner, collecting a capella recordings from professional singers and singing students, which permits to get pairs of recordings. It is composed of two opera role types, \textit{dan} and \textit{laosheng}, and includes 20 different reference melodic lines of each role type with corresponding syllable-level annotations. The dataset has initially been recorded to evaluate the singing quality of the students compared to professional singers. This implies that the recordings sometimes contain mistakes in the lyrics, and breaks in between sentences. For each reference line, we count between 1 and 10 target versions that serve as target in our experiments.

\section{Experiments and Discussion}\label{sec:experiments}

\subsection{Acoustic Model Training}\label{sec:AM_training}

For our experiments, we train two distinct acoustic models, based on different training subsets of DALI dataset\footnote{We also tried to train a third acoustic model only on Italian data, which is the target language of the ``Don Giovanni'' opera. However, all alignments diverged. We believe that this is mainly due to the low amount of training Italian data, 10.8~h, in the DALI dataset, which is significantly lower than the other proposed languages.}. The first model, \textit{5lang}, includes songs from five languages, namely English, German, French, Spanish, and Italian. The second model, \textit{english}, uses only English data, the most represented language in the dataset. The different train and validation splits were made publicly available by \cite{vaglio2020multilingual}\footnote{\url{https://github.com/deezer/MultilingualLyricsToAudioAlignment}} and are described in Table~\ref{tab:training_dataset}. Each acoustic model is trained with the CTC loss, a learning rate of $10^{-4}$, and the ADAM optimizer.

\begin{table}[t]
\small
 \begin{center}
 \scalebox{1}{ 
 \begin{tabular}{@{}lrr@{}}
  \toprule
  \textbf{Dataset} &  \textbf{Train songs (duration)} & \textbf{Valid. songs (duration)}\\
  \midrule
  \textbf{5lang} & 4027\ \ \ (259.4~h) & 149\ \ \ (9.4~h)\\
  \textbf{english} & 3257\ \ \ (210.8~h) & 39\ \ \  (2.6~h)\\
  \bottomrule
 \end{tabular}}
\end{center}
 \caption{Acoustic Model training and validation datasets.}
 \label{tab:training_dataset}
\end{table}

\subsection{Evaluation Metrics}\label{sec:metrics}

As mentioned above, we evaluate our lyrics trackers by quantifying the precision of the alignment between target and reference that they produce. The granularity of our ground truth annotations is at the bar level for Don Giovanni and at the syllable level for the Jingju dataset. We use the standard evaluation metrics from the field of score following \cite{cont2007evaluation}. For each alignment, we report the mean tracking error, in seconds, between timestamp annotations and times detected by our aligner. We also report the proportion of annotations that are detected with an error less than 1s.

\subsection{Lyrics Tracking in Opera}\label{sec:opera_lyrics}

To evaluate our system, we compare its performance with other following techniques working in real-time. To this end, we select the State-Of-The-Art (SOTA) live opera tracker that has recently proved to be robust to track, from beginning to end, complete live ``Don Giovanni'' performances \cite{brazier2020addressing, brazier2020towards}. The opera tracker applies an OLTW algorithm to align reference and live target audio. Instead of using posteriograms as input to the OLTW algorithm, it takes audio features that are directly computed from the audio recordings. For the study, we compute two types of features that will serve as tracking baselines and inputs to our alignment algorithm. The first feature, \textit{baseline}, has been inspired from music tracking systems performing on orchestral performances \cite{gadermaier2019study}. The feature calculates 120~MFCCs, but discards the first 20, and is computed at a sampling rate of 44.1~kHz, a window size of 20~ms, and a hop size of 10~ms. The second feature, \textit{recitative}, was designed specifically with recitatives in mind \cite{brazier2020addressing}. It was tuned to perform best on the recitative subset of the Fischer recording and was shown to generalize well to the recitative subset of Manacorda. The feature is composed of 25~MFCCs extracted from Linear Predictive Coefficients (LPC) that aim at extracting the phoneme information from the audio. It is computed at a sampling rate of 1500~Hz, with the same previous window size and hop size.

\begin{table}[t]
\aboverulesep=0ex
\small
 \begin{center}
 \scalebox{1}{ 
 \begin{tabular}{@{}l|llrc@{}}
  \toprule
  \textbf{Opera} &  \textbf{Name} & \textbf{Feature} & \textbf{Mean(ms)} & $\mathbf{\leq 1s}$\\
  \midrule
  \textbf{DG} & Targ\_Fischer & baseline      & 1,915 & 66.1\%\\
              &               & recitative    & 955  & 76.5\%\\
              &               & 5lang         & 846  & 80.5\%\\
              &               & english       & \textbf{818}  & \textbf{80.5\%}\\
  \cmidrule{2-5}
              & Targ\_Manacorda & baseline    & 1,503 & 62.0\%\\
              &                 & recitative  & 1,023 & 69.7\%\\
              &                 & 5lang       & \textbf{824} & \textbf{77.6\%}\\
              &                 & english     & 963 & 76.1\%\\
  \midrule
  \textbf{Jingju} & Targ\_Jingju & baseline   & 2,943 & 61.5\%\\
                  &              & recitative & 3,878 & 60.0\%\\
                  &              & 5lang      & 964 & 87.5\%\\
                  &              & english    & \textbf{810} & \textbf{89.0\%}\\

  \bottomrule
 \end{tabular}}
\end{center}
 \caption{Tracking error on Don Giovanni (DG) and Jingju opera sub-datasets}

 \label{tab:results}
\end{table}

The results are given in Table~\ref{tab:results}. Looking at Don Giovanni / Fischer, we see that both new models, using our \textit{5lang} and \textit{english} acoustic models, outperform the SOTA opera tracker based on \textit{baseline} and \textit{recitative} features, the latter of which had been optimized specifically on this dataset. The mean error has been reduced by at least 100~ms and 80.5\% of the bars now show an error below 1s. A similar picture emerges with DG / Manacorda, where the mean error goes to 824~ms for the best \textit{5lang} model, and where the 1s threshold improves by 8 percentage points, relative to the recitative feature tracker. It is also important to note that the results of \textit{recitative} and \textit{baseline} were obtained in combination with a dedicated silence detector which halts the tracking process when there are obvious pauses. The two new trackers simply use the posteriograms generated by the acoustic models and still improve tracking accuracy.

Secondly, our two models \textit{5lang} and \textit{english} also perform best, by a large margin, on the Jingju opera sub-dataset. The \textit{baseline} and \textit{recitative} features were designed to extract pitch contours, focusing at different parts of the frequency range. They turn out to be inefficient at tracking Beijing Opera a capella recordings. On that task, the \textit{english} model achieves the best performance, with a tracking accuracy of 810~ms and 89.0\% of the syllables being detected below 1s of error.

\subsection{Robustness to Different Languages}\label{sec:robustness_languages}

Comparing the results, we see that alignment accuracies are very similar across the two corpora, even though they contain singing signals in two very different languages, Italian and Chinese. The two acoustic models were not trained on Chinese recordings,\footnote{and indeed, Don Giovanni's language -- Italian -- was also not represented in the \textit{english} acoustic model's training data.} a language that includes new phonemes that do not appear in the phoneme set built from the five European languages present in DALI. Using phoneme posteriograms as a joint intermediate representation of the lyrics and of the live input thus seems to be a remarkably robust choice for multilingual tracking.

Finally, the best results seem to alternate between the \textit{5lang} and \textit{english} acoustic models in the different sub-datasets. Based on the conclusions of \cite{vaglio2020multilingual}, we expected the \textit{5lang} model to perform best. However, and especially on the Jingju tracking experiment, \textit{english} yields sometimes better performance. This may be explained by the fact that the mapping into the European phoneme set is by construction incorrect, since the Chinese language is not in the training dataset. We expect that training our acoustic model with additional Chinese data would boost the performance.

\section{Conclusion}\label{sec:conclusion}

We have presented an on-line audio-to-lyrics alignment method that is capable of operating in a real-time scenario. It involves an acoustic model, built from a ResNet architecture, that classifies each audio frame into a vector representing the probability distribution over a predefined set of phonemes, with a delay of 280~ms. In a second step, a real-time capable alignment algorithm (On-Line Time Warping) aligns the emerging sequence of vectors to a posteriogram matrix that has been extracted beforehand from a reference performance of the same work, via our acoustic model. In experiments, we showed that our method is robust and reasonably precise in tracking the lyrics in a musical genre where the sung lyrics are known to be hard to understand, and a genre that was not part of the acoustic model training dataset. Additionally, we also showed robustness across languages, even if these are not included in the acoustic model training data. Our results suggest that it might be fruitful to investigate combinations of our system with existing music trackers in the more general task of opera score following.

Moreover, even if our study focused on the specific genre of opera (and on two very specific subsets of it), the method should be directly applicable to other music genres and other languages. The acoustic model was trained on Western musical genres and consequently, we expect it to work even better on those genres. As future work, we plan to evaluate our system on available Western music datasets containing pairs of recordings, such as Covers80 \cite{ellis2007identifying}, where lyrics are not necessarily identical and where song structures may differ, and to use offline lyrics alignment systems to obtain reference annotations.

\section{Acknowledgments}\label{sec:acknowledgments}

The research is supported by the European Union under the EU's Horizon 2020 research and innovation programme, Marie Sk\l{}odowska-Curie grant agreement No.~765068
(``MIP-Frontiers").
The LIT AI Lab is supported by the Federal State of Upper Austria.
Thanks to Andrea Vaglio, Emir Demirel, and Khaled Koutini for our interesting discussions about this work.

\bibliography{ISMIRtemplate}

\end{document}